# The challenges and realities of retailing in a COVID-19 world: Identifying trending and Vital During Crisis keywords during Covid-19 using Machine Learning *(Austria as a case study)*

**Reda Mastouri**

Research proposal for the graduate assistantship (GA) in Data Sciences

at the Saint Peter's University

9/4/2020

# Table of contents

# List of figures and Tables

# List of acronyms and abbreviations

API            Application programming interface is a set of functions and procedures allowing the creation of applications that access the features or data of an operating system, application, or other service.

COVID19        COVID-19 (coronavirus), known in full as SARS-CoV-2 ("the virus") was detected in Wuhan, Hubei Province, China in Dec 2019. Real time worldwide impact can be viewed via this link: https://covid19.who.int/

DSI            Downstream Impact: DSI is an econometric data-driven model that estimates the causal medium to long-term effect of a customer action on the customer's subsequent purchasing behavior on a retail marketplace. Today, DSI is used to study semi-permanent customer actions (that tend to stick to the customer). Any seasonal/holiday events may not produce immediate revenue, but they often signal a shift in long-term purchasing behavior by reinforcing customers' loyalty, making digitized supplay chaim marketplaces more accessible, and/or providing new marketing channels. It answers questions like **"What change in purchases over the next X days (typically 365) do we predict for this customer if he/she were to take a particular action today, rather than at some point in the future (or not at all)?".** DSI metrics are established as the standardized estimate for long-term value of customer actions and are being integrated into Amazon's marketing systems as a key input to optimize content prioritization.

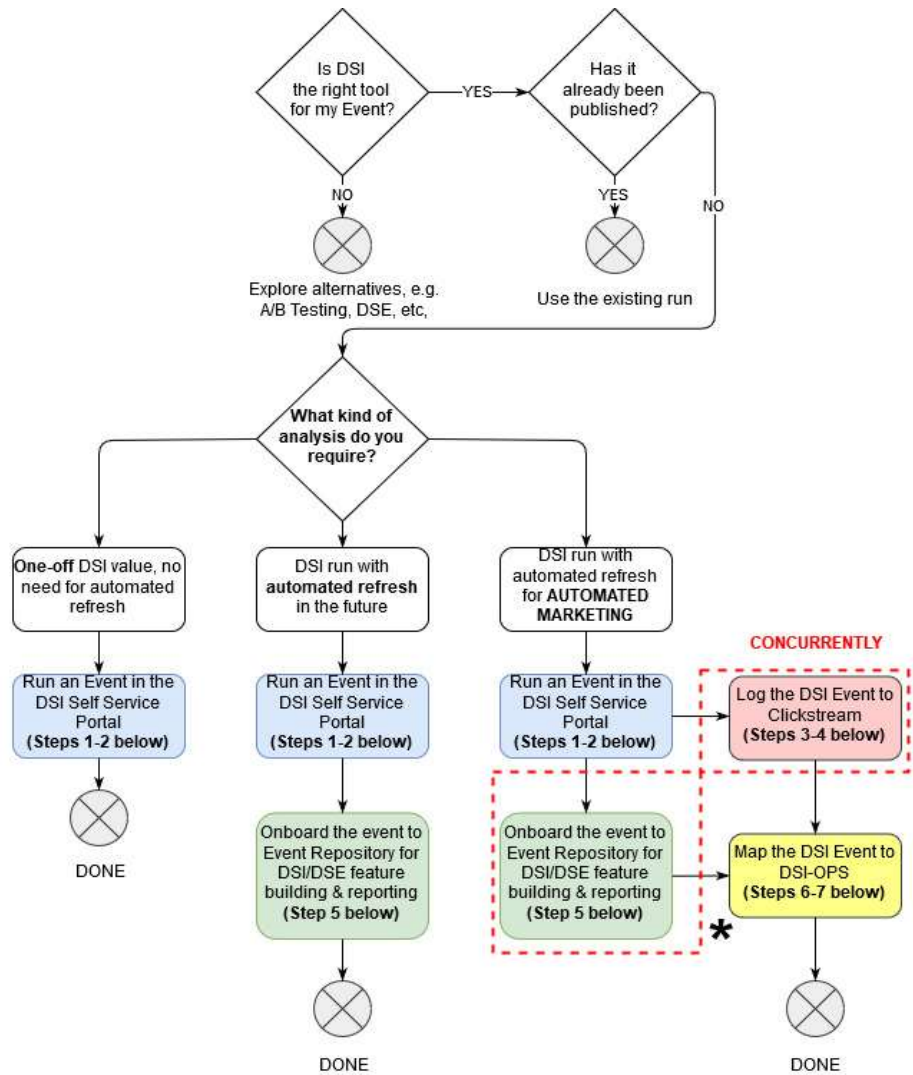

**Figure IV.1: DSI Model**

ETS             is broad family forecast methods on top of **Exponential smoothing state space.**

Random Forest   is one of the ensemble learning method could be used for classification. consists of a large number of individual decision trees that operate as an ensemble. It utilizes the wisdom of crowds.

Forecasting     The forecast is a system calculation that predicts unconstrained customer demand

Unconstrained = Units we expect to sell with infinite inventory

Forecasts in quantiles (P50, P90, etc.) that express uncertainty
- P50: units that will cover actual demand for 50% of times
- P90: units that will cover actual demand for 90% of times

Keyword

Essentiality    The essentiality of the keyword is calculated based upon a background running algorithm that generated few score to the respective keyword if applicable

Lambda          AWS Lambda is an event-driven, serverless computing platform provided by Amazon as a part of Amazon Web Services. It is a computing service that runs code in response to events and automatically manages the computing resources required by that code. It was introduced in November 2014.

LSTM        Long short-term memory is an artificial recurrent neural network architecture used in the field of deep learning. Unlike standard feedforward neural networks, LSTM has feedback connections. It can not only process single data points, but also entire sequences of data.

Prophet     is a procedure for forecasting time series data based on an additive model where non-linear trends are fit with yearly, weekly, and daily seasonality, plus holiday effects. It is open source software released by Facebook's Core Data Science team.

RedShift    Amazon Redshift is a data warehouse product which forms part of the larger cloud-computing platform Amazon Web Services. The name means to shift away from Oracle, red being an allusion to Oracle, whose corporate color is red and is informally referred to as "Big Red."

RNN        Recurrent neural network is a recurrent neural network is a class of artificial neural networks where connections between nodes form a directed graph along a temporal sequence. This allows it to exhibit temporal dynamic behavior.

S3           Amazon S3 or Amazon Simple Storage Service is a service offered by Amazon Web Services that provides object storage through a web service interface. Amazon S3 uses the same scalable storage infrastructure that Amazon.com uses to run its global e-commerce network.

SARIMA    A **seasonal** autoregressive integrated moving average (**SARIMA**) model is one step different from an **ARIMA** model based on the concept of seasonal trends. In many time series data, frequent seasonal effects come into play. Take for example the average temperature measured in a location with four seasons

Seasoanlity  Seasonality is a pattern of demand that repeats every year for certain products. E.g. Swimsuits during the summer or cold-weather boots during winter.

SQL        for Structured Query Language. SQL is used to communicate with a database. According to ANSI (American National Standards Institute), it is the standard language for relational database management systems

Trailing    typically refers to a certain time period up until the present. For example, a 12-month trailing period would refer to the last 12 months up until this month. A 12-month trailing average for a company's income would be the average monthly income over the last 12 months. Taking an average like this can help smooth out fluctuating or cyclical data series. A trailing average may also be referred to as a moving average. Gather your data and arrange it in chronological order with the time periods noted (for example, January income, February income and so on).

TPV       Third party verification (TPV) is a process of getting an independent party to confirm that the customer is actually requesting a change or ordering a new service or product.

TTM       The **transtheoretical model** of behavior change is an integrative theory of therapy that assesses an individual's readiness to act on a new healthier behavior, and provides strategies, or processes of change to guide the individual.

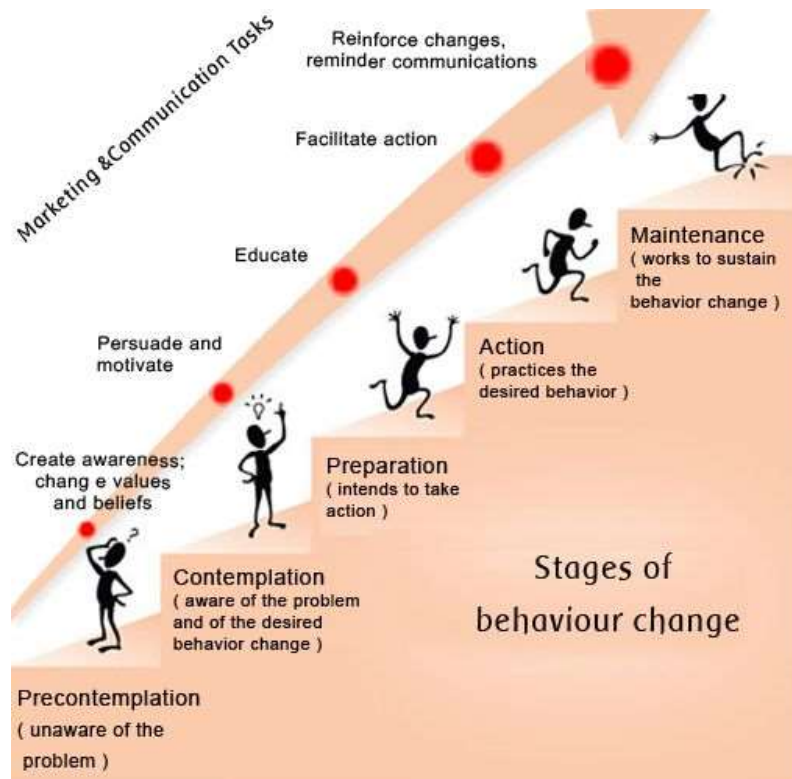

**Figure IV.2: trans-theoretical model**

TTM       **Trailing 12 months** (TTM) is the term for the data from the past **12** consecutive **months** used for reporting financial figures. A company's **trailing 12 months** represent its financial performance for a **12-month** period; it does not typically represent a fiscal-year ending period.

TPV       **Total Payment** Volume or **TPV** is the **value** of **payments**, net of **payment** reversals, successfully completed through **Payments** Platform, excluding transactions processed through our gateway products.

WHO     World Health Organization

XGBoost  is an implementation of gradient boosted decision trees designed for speed and performance. It is an optimized distributed gradient boosting library designed to be highly **efficient**, **flexible** and **portable**.

# 1.    INTRODUCTION

Covid19 has caused a huge change in consumer's shopping behaviours. With the uncertainty that prevailed and confusion that got created due to different advisories and information from different sources people preferred to be over prepared than underprepared. There was a phase of panic buying where people stocked up their houses with all essential items in fear of a prolonged lockdown. One of the main questions that people faced was that "Is online shopping safe?". There were several reports available suggesting that the risk of spreading of virus through ordered products is very little considering the increased delivery time and also the fact that shipping conditions make a tough environment for the virus. The WHO also released a statement in its favour saying - "The likelihood of an infected person contaminating commercial goods is low and the risk of catching the virus that causes COVID-19 from a package that has been moved, travelled, and exposed to different conditions and temperature is also low."

Some examples of reports suggesting surge in e-commerce business as an impact of COVID19 –

- U.S. e-commerce jumped 49% in April, compared to the baseline period in early March driven by online groceries and electronics - LINK

- E-COMMERCE SALES Increased BY 52% LINK

- The increase in ecommerce spending since pre-pandemic days appears to have settled in at about the 50% range despite the opening up of brick-and-mortar - LINK

We need models to navigate the supply chain with greater precision and forecasting. Forecasting is a critical input to our supply chain as it impacts buying decisions.

| Category | Method |
|----------|--------|
| Time-series | Naïve, Moving Averages (four models 2,3,4,7), SES, ETS, ARIMA, Theta, TBATS, ANN_AR, G&M (1985)-Damped trend (Gardner & McKenzie, 1985), Holt - Trend, ns-HW (non-seasonal Holt-Winters), ARFIMA, GARCH(1,1) (six models, wih: GED, SGED, NORM, SNORM, STD, SSTD), ARIMAx, Naïve-d with drift (ten models with step of 0.1 for the drift) |
| Machine Learning | Multiple linear regression (MLR), Ridge regression, Decision Trees (DT), Random forest (RF), Neural Network (NN), Support vector machine (SVM). |
| Deep Learning | Long-Short Term Memory networks (LSTM) |
| Others | Splines, Sigmoid, Partial Curve Nearest Neighbor methods (PC__NN), Multivariate Clustering based Partial Curve Nearest Neighbor methods (CPC__NN) |

**Table 1.1: Forecasting methods**

From global pandemics to geopolitical turmoil, leaders in logistics, product allocation, procurement and operations are facing increasing difficulty with safeguarding their organizations against supply chain vulnerabilities. It is recommended to opt for forecasting against trending based benchmark because auditing a future forecast puts more focus on seasonality. The forecasting models provide with end-to-end, real time oversight of the entire supply chain, while utilizing predictive analytics and artificial intelligence to identify potential disruptions before they occur. By combining internal and external data points, coming up with an AI-enabled modelling engine can greatly reduce risk by helping retail companies proactively respond to supply and demand variability. This research paper puts focus on creating an ingenious way to tackle the impact of COVID19 on Supply chain, product allocation, trending and seasonality.

*Key words: Supply chain, covid-19, forecasting, coronavirus, manufacturing, seasonality, trending, retail.*

## 2. OVERVIEW OR BACKGROUND

The impact of COVID-19 on companies is evolving rapidly and its future effects are uncertain. While the novel COVID-19 spreads globally, the retail industry braces for the outbreak's fallout.

In Austria, officially the Republic of Austria, the COVID-19 crisis is strongly affecting state and municipal budgets. A 7%-12% drop in state tax revenues is forecasted according to oced.org. Based on statista.com insights, retail in Austria demonstrated the most dramatic decline on March 29, 2020 at 87%. Most recently, supply chain declined as well by 43 percent in the week of May 2, 2020. In August 2020, 33.2% more unemployed persons were registered in Austria than in the previous month, totally amounted to 371,893 unemployed citizens by national definition. Robert Holzmann, Austrian National Bank (OeNB) Governor, said the central bank now forecasts an 8% drop in Austria's gross domestic product (GDP). It had previously foreseen a 3.2% drop in a "moderate" pandemic scenario, estimating that each week of lockdown cuts annual GDP by more than $2 billion.

Retail Sales in Austria is expected to be 0.20% by the end of the Q3, according to Trading Economics global macro models and analysts' expectations. Looking forward, we estimate Retail Sales in Austria to stand at 0.20 in 12 months' time. In the long-term, the Austria Retail Sales is projected to trend around 0.20% in 2021 and 0.10 percent in 2022, according to our econometric models.

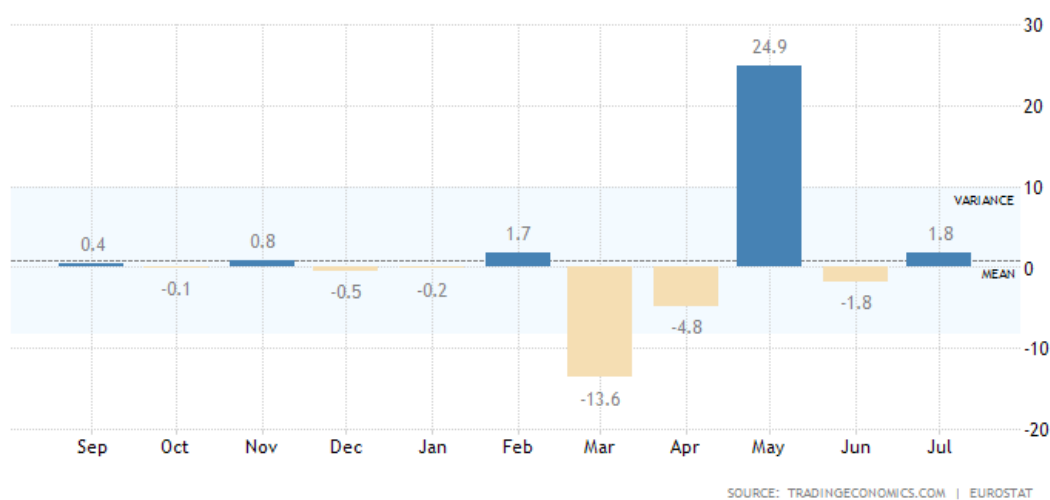

**Figure 2.1: Retail Sales in Austria increased 1.80 percent in July of 2020 over the previous month.**

"Retail has hung a closed sign on the door literally and metaphorically," Neil Saunders, managing director of GlobalData Retail. "This is the most catastrophic crisis that retail has faced – worse than the financial crisis in 2008, worse than 9/11. Almost overnight, the retail economy shifted from being about things people want to things that they need."

| Summary of impact of the Covid-19 on Retail in Austria | |
|---|---|
| **Status – Has any special status been introduced?** | **State of emergency and curfew:** Based on the Law on Measures against COVID-19, the Minister of Social Affairs issued ordinances pursuant to which persons in Austria are not allowed to leave their homes except for the following reasons: <br><br> • necessary journeys to work; <br><br> • ensuring their basic needs or the basic needs of people in need (such as essential trips to the doctor, food shops, post office, bank, pharmacy, gas station etc.); <br><br> • visits to funerals or marriages of close relatives; and <br><br> • short outdoor activities (e.g. walks, runs, etc.). Also in these cases, there is a general obligation to keep at least a one-meter distance from other persons (the minimum distance does not apply between persons living together in the same household) and to wear protective masks (in retail areas of shops and in public transportation). <br><br> The measures are valid until April 30, 2020, and can be prolonged / amended. Noncompliance may trigger fines of up to €600. |
| **Which retail units are Open** | The following retail and service units are expressly permitted to operate and to be entered: <br><br> • public pharmacies and hospitals; <br><br> • retail supermarkets, local food retail stores; <br><br> • petrol stations; <br><br> • restaurants (only delivery or take-out); <br><br> • veterinary clinics; <br><br> • hardware stores; |
| | <ul><li>maintenance and repair workshops for vehicles;</li><li>bank, and post offices;</li><li>legal professions;</li><li>delivery services;</li><li>public transportations;</li><li>waste management companies;</li><li>tobacco and newspaper shops; and</li><li>agricultural trade companies.</li></ul>In addition, all other retail and service units engaged in the area of sale, manufacture and repair of goods, which are not expressly mentioned in the exception list, and which have a maximum customer area of 400 sqm, are permitted to open if certain additional conditions are met (protective masks, minimum distance and at least 20 sqm of the customer area available to each customer).<br><br>These measures are valid until May 3, 2020. Non-compliance may trigger fines of up to €30,000. |
| **Which retail units are Closed** | The following units are required to be closed<ul><li>All leisure shops, shopping centres, hairdressers, establishments open to the public, offering cultural, social, festive, sporting and recreational activities, as well as playgrounds (unless an opening exception applies).</li></ul>Hotels, restaurants, cafés and bars (with a possibility to provide delivery or take-out services). |
| **Leases: Have special laws related to COVID-19 been implemented** | So far, only special laws regarding residential premises have been implemented. |
| **Employment Packages** | **Coronavirus short-term work:** Under this model, the employer pays a part-time salary and, in addition, the short-time work subsidies which he will be refunded by the labour market service. A reduction of working time by up to 100%, and thus a complete release of employees, is possible for certain periods. However, the planned working time must be between at least 10% and 90% within an initial averaging period of three months. |
| **Tax Rebates** | **Tax regulations:** Reductions or non-assessment regarding advance payments of income or corporate income tax (including on interest) are possible in case of liquidity shortages due to a COVID-19 infection. Furthermore, companies may request deferrals and payment in instalments, as well as a no assessment of deferral interest.<br><br>**Support measures by social security institutions**: The Social Security for the Self Employed (Sozialversicherung der Selbständigen) provides support in case of financial losses due to the Coronavirus pandemic by deferral of contributions, agreement on instalments, the reduction of the provisional contribution base, as well as full or partial non-determination of interest for late payment.<br><br>Likewise, the Austrian Health Insurance (Österreichische Gesundheitskasse) provides support to affected employers.<br><br>**Stamp duties:** A comprehensive exemption from stamp duties in connection with incidences following from COVID-19 measures has been enacted |

**Table 2.1: Summary of impact of the Covid-19 on Retail in Austria**

According to GCP The long-tail impacts of the COVID-19 pandemic will continue to be felt in the retail industry through this year and into the next. Many retailers have been forced to reduce their physical footprints or change the way they operate, in addition to exploring new ways of delivering their products into the hands of customers, with more changes likely to come.

COVID-19 is a crisis and a threat, but it also presents opportunities to leaders who understand the short-term prospects of their industries. Executives and operations in Austria and the other countries have a unique opportunity to start deploying AI-driven solutions focusing more on emerging market trends and providing actionable insights to help them identify market opportunities and develop effective strategies to optimize their market positions with the help of analytics and data sciences.

## 3.    RESEARCH FOCUS

We need a scalable model that uses historical demand and vendor availability signals. This section explains our forecasting model building framework and auditing mechanisms.

### 3.a. Seasonality:

The job is to set out to create the foundation for a scalable forecasting model that addresses seasonality. It is important to build a model in reusable components to expedite later expansion to additional use cases. The framework is to (1) normalize historical demand, (2) identify allocated products to forecast, (3) generate a point estimate, (4) create a forecast distribution, and (5) back-test and minimize risk.

*Example*: The performance of the top trending keywords on Amazon [last 15 weeks] and Google search [last 5 years] can be compared in this section.

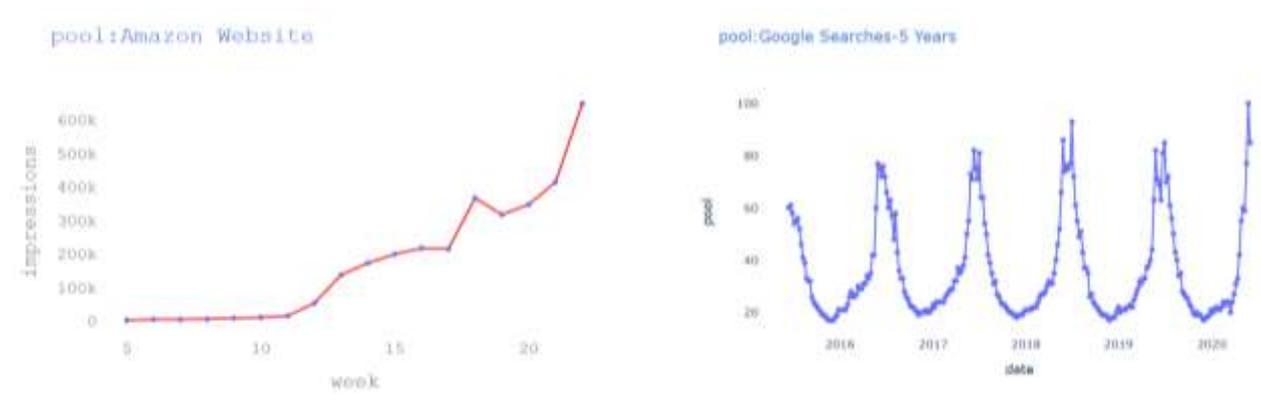

**Figure 3.1: Amazon searches and Google trend for seasonality**
**Source: Google Analytics 08-2020**

To understand about all the products that fall under the keyword "pool" in Amazon right click on Amazon website word in pool: Amazon website and open the link. It will take us to all the potential products that can be capitalized at the moment by consumers.

**What problem does this solve?**  Seasonality enables users to easily identify seasonal demands through a consistent methodology, empowering them to take business actions accordingly. For example, Forecasting can audit forecast accuracy by seasonal groups to identify defects, and most concerned retailers can take markdown decisions differently for seasonal products that are entering vs. exiting their peak season. Ultimately, it is still a big question mark about how seasonality is quantified!

### 3.a. Audit and Tracking:

In parallel with creating these initial models, it is imperative to build auditing and tracking mechanisms to understand where to double down on overrides and where to take our hands off the wheel.

As forecast overrides are static and the system forecast is dynamic, there will be a need of building standardized rules to determine when to remove forecast overrides. We revert to the forecasting system when the system catches up to our override, when allocated products are moved to the neural network models, and when new products are first received. These rules are re-applied weekly.

It is indispensable to measure the intervention performance by comparing the aggregate 90% weighted error between the overrides and the system. In parallel, auditing the forecasting system to proactively identify defects remains critical. In the next step, we need to compare the aggregate forecast for different seasonality groups (Fall/Winter and Spring/Summer etc.) against a benchmark calculated using the trailing 12-week year over year growth trend.

### 3.1    RESEARCH PROBLEM

The following are the challenges we are expecting during model building:

a) No initial starting point to find risky/trending product in March 1st week i.e. Week 1 of Covid19

b) Limitations with Kaggle dataset, search keyword, Catalog index keywords and manual keywords- Since Kaggle was limiting its priority list to only top sellers and top brands, it is insufficient to prevent embarrassing price increases on all tends

c) Keyword ratio bump in customer searches provided a lot of False positives.

d) Psychological differences in needs vs Wants during Covid19 across individuals

e) Absence of tagged keywords to ascertain degree of essentiality during Covid19.

f) Model to finding similar allocated products based on sentence embedding- Since it is a recent development, documentation is limited.

g) Scaling the analysis to find essential keywords across multiple languages.

## 3.2     RESEARCH QUESTION

3.2.a. **Strengthen the strategic view of supply**

- How responsive is my supply chain? How quickly can I make decisions?
- Do I understand my supplier reliance? Where are the critical risks?
- How did the change in customer shopping behaviour affect latency?
- Where are my key suppliers located? Would diversifying geographies reduce risk?
- Do I have an inventory strategy? Does it consider the risk of future disruption?

3.2.b **Invest in digital supply chain technologies**

- Do I have end-to-end visibility of my supply chain? Where are the risks and gaps?
- Can I predict demand? Where could I benefit from more powerful demand prediction?
- Do I have robust scenario planning in place? Could tools such as digital twins help me understand the potential impact on the whole business from shocks?

## 3.3     RESEARCH AIM

The goal of this exercise is to leverage a science-based and data-driven methodology to recommend a baseline for setting the 2021 treading and seasonality goals (Payment Volume/Buyers). 2020 has been an unusual year with COVID pandemic affecting the retail businesses in different manner in different countries. Here the forecasts are built using two different scenarios one with COVID period data included and one without. This will give an understanding of the future impact of COVID on the supply chain as a whole.

3.3.a. **Without Covid19 impact**

Due to COVID19 there has been a huge impact on many companies' metrics and buyers which varies across countries. This impact is an anomaly driven by an event so the idea here is to look at the metric forecasts assuming that this event didn't occur.

3.3.a. **With Covid19 impact**

COVID19 impact varies across regions and merchants. This will also impact the future metrics hence the purpose of these experiments is to build a forecasting model for rest of 2020 and 2021 to forecast the metrics (Total Payment Value and Buyers) including the COVID19 impact.

# 4.    RESEARCH METHODOLOGY

## 4.1. Without Covid19 impact- Method

In this scenario 2020 data cannot be included as COVID19 started in March and we do not want to include that impact. So data from 2015-2019 will be used for forecasting 2020-2021.

**Method Summary**

a) Take data from 2015-2019 (Use the companies' data/ Kaggle).

b) Run log transformation

c) Build a peak flag. This flag will have 1 for Black Friday and Cyber Monday weeks and 0 for rest of the year. Name a given marketplace event having a huge peak during these two weeks driven by this event hence this flag is to point out the event time.

d) Optimize parameters using Auto ARIMA

e) Using the optimized parameters run SARIMAX to forecast the 2020

f) Use the data till 2020 to again train and forecast 2021

| | 2015 | 2016 | 2017 | 2018 | 2019 | 2020 | 2021 |
|---|---|---|---|---|---|---|---|
| New (USD) | - | - | - | - | - | - | - |
| Reengaged | - | - | - | - | - | - | - |
| **TPV** (USD) | - | - | - | - | - | - | - |
| **TTM** (USD) | - | - | - | - | - | - | - |
| | | | | | | | |
| YOY | 2015 | 2016 | 2017 | 2018 | 2019 | 2020 | 2021 |
| New (USD) | - | - | - | - | - | - | - |
| Reengaged | - | - | - | - | - | - | - |
| **TPV** (USD) | - | - | - | - | - | - | - |
| **TTM** (USD) | - | - | - | - | - | - | - |
| | | | | | | | |
| Average | 2015 | 2016 | 2017 | 2018 | 2019 | 2020 | 2021 |
| New (%) | - | - | - | - | - | - | - |
| Reengaged (%) | - | - | - | - | - | - | - |

**Table 4.1: Summary layout**

## 4.2. With Covid19 impact- Method

Below are some of the proposed experiments to forecast buyer and TPV metric forecasting for year 2021 incorporating the COVID impact

**Univariate Time Series Forecasting Experiments**
Only the metric data is used for forecasting, no other dependent variables will be used. This model would be based on the past trend and seasonality of the metric.

**Experiment 1**
Add the COVID event as a flag in the training data where the COVID period is marked as 1 and rest is 0.
For forecasting future scenario based assumptions will be made for COVID19 impact.

**Scenario 1 : COVID impact peak has already been observed and now there will be minor impact in 2020 and no impact in 2021**

1. Take data from 2015-2021 (Use the companies' data/ Kaggle).

2. Run log transformation

g) Build a peak flag. This flag will have 1 for Black Friday and Cyber Monday weeks and 0 for rest of the year. Name a given marketplace event having a huge peak during these two weeks driven by this event hence this flag is to point out the event time.

3. Build a COVID flag. This flag will have 1 for 2020 peak weeks, -1 in those weeks in 2021 and 0 for rest of the years

4. Optimize parameters using Auto ARIMA

5. Using the optimized parameters run SARIMAX to forecast next 78 weeks which is rest of 2020 and 2021

**Multivariate Time Series Forecasting Experiments**

The metrics are dependent on other variables for example number of buyers are dependent on the merchant growth, so a multivariate time series model can take all the dependent parameters while forecasting the metric. These models are based on the past trends and seasonality of the metric as well as the relationship with other variables.

**Experiment 2**
From the COVID impact analysis some of the major factors affecting TPV and buyers were:

1. Varying merchant performance in different categories

2. Varying merchant performance in different TPV buckets

3. Gaining New Buyers

The following **variables** can be used in the model based on the above observations from the analysis

1. No. of transacting merchant every week (Number of weekly transacting merchant increases with increasing TPV)

2. Divide TPV in buckets (High, Mid, Low) and then put number of merchants in each bucket every week.

3. Number of merchants bringing New Buyers

4. Group categories based on COVID impact (Major Growth, Mid Growth, Major Drop, Mid Drop, no impact), add number of transacting merchants in each group.

5. Alexa ranking (May be for New merchants, will have to check data)

6. Anomaly period flag

7. Holiday

4.3. **Multivariate Models**

1. Version of **ARMAX**. (Vector Auto Regression Moving-Average with Exogenous Repressors): VARMAX is a generalisation/multivariate version of ARMAX.

2. **LSTM** (Long Short Term Memory): LSTM is an artificial recurrent neural network (RNN) which unlike a standard feedforward neural network also has a feedback connection. It can process sequence of data and is popularly used in time series forecasting. (Reference : https://colah.github.io/posts/2015-08-Understanding-LSTMs/)

**Experiment 2**

1. **VARMAX**
   **Data:** Variables - "No. of Transacting Merchants" and "TPV"
   **Method:** Remove seasonality by differencing by 52, convert to log and then run VARMAX
   **Result:** Flat forecasts or overshooting with the straight line

2. **LSTM**
   **Data:** Variables - "No. of Transacting Merchants" and "TPV"
   **Method:** Simple sequential LSTM with 2 LSTM layers and dense layer
   **Result**: Model could experience convergence issue (Data is not enough to train). Difficult to make validation set with history as 2 years.

## 4.1 QUALITATIVE RESEARCH

Below is the architecture we propose based on AWS technologies:

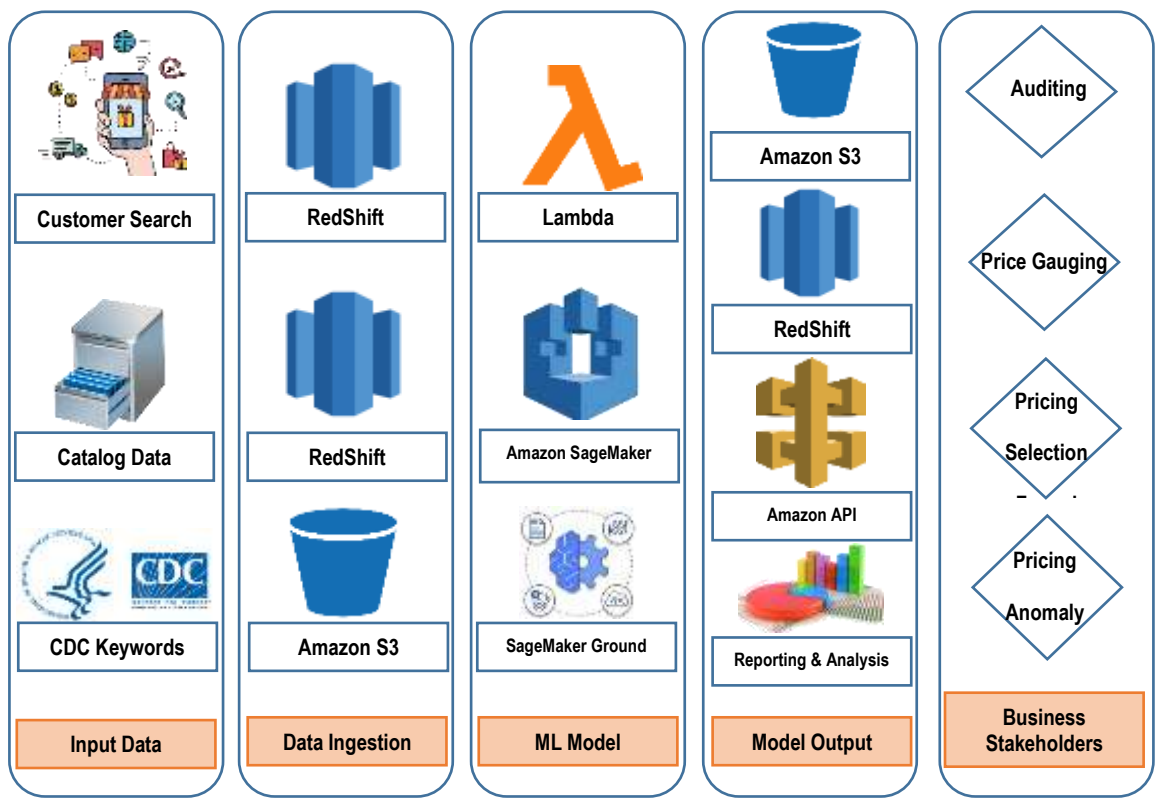

**Figure 4.1: Proposed architecture**

## 4.2    QUANTITATIVE RESEARCH

Dynamically capture changing customer behavioural patterns and increase the coverage of Covid-19 demands limited by heuristic keywords like 'Face Mask', 'Hand Sanitizer' etc. Also, categorize a given product based on the customers' Needs Vs Wants in an unsupervised manner and predict its degree of essentiality during the crisis.

## 4.3    DATA COLLECTION

Our model uses customer search data, Catalog index data for the targeted products, item descriptions on detail page on the retailer database for ground truth enrichment using AI.

Process flowchart to generate scores per demands/needs by using an ensemble of modelling techniques ranging from NLP based word embedding to Topic Modelling to manual ground truth generation

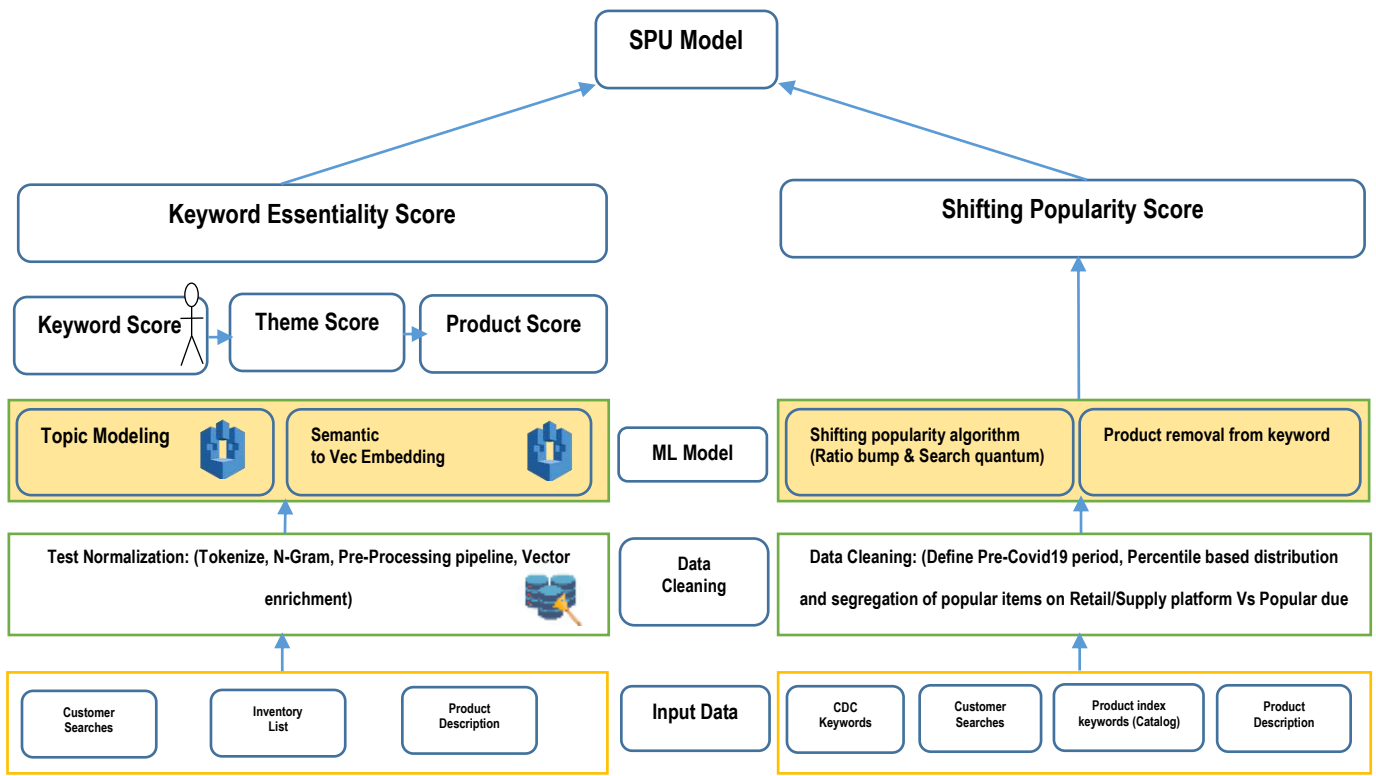

**Figure 4.2: Workflow for Trending and Essential Keywords**

One of the techniques that can be used are from multiple unrelated disciplines such as Topic Modelling & Word Embedding from Machine Learning, Maslow's hierarchy of needs from philosophy and qualitative word tagging from Netflix Recommendation engine to identify the degree of essentiality (0-1 scale) of an item during crisis. To ensure model's correctness and swift applicability, it is recommended to tag ~2000 English words as ground truth input to NLP models and use their translations across multiple languages.

Here is a list of syndicated data we can use for benchmarking:

| Data Source | What data do they have? |
|---|---|
| MINTEL | Mintel's library of Food and Foodservice reports has approximately 50 titles (ranging from "Snacking Motivations and Attitudes" to "Convenience Store Foodservice.") Each report is approximately 50-80 doc pages in length and includes detailed consumer research, sales data and market information, along with analysis, context and commentary covering the past, present and future of different food/beverage topics.<br>Reading a Mintel report (or two) is a good first step as you seek answers to any questions. While it won't likely provide all the answers we may seek, it will provide a good overview of what's happening and sharpen our focus as we move forward seeking additional data and insights. |
| HARTMAN RETAINER SERVICE | The Hartman Group is a foremost authority on consumption behaviours, consumer motivations and cultural trends in the food and beverage category. Whole Foods currently considers The Hartman Group as their gold-standard syndicated data and insight partner.<br><br>The Hartman Retainer Services (HRS) is a customized subscription service that provides access to all of The Hartman Group's non-proprietary reports and publications. Via this partnership, we can access a wealth of syndicated studies, trend reports, podcasts and white papers on food/beverage, health & wellness and customer food shopping topics. In addition, you have access to Hartman's annual Compass eating occasions database.<br>As part of our service agreement, HMR has a bank of consulting hours for specific questions we may have. |
| NIELSEN | Nielsen has an interesting database called Scanner and Panel data. Nielsen is the definitive source for: market share and performance drivers; product innovation, trends and seasonality; share of wallet and leakage; and quantifying size of prize. While this data is available as self-service reports for the users, we have a robust mechanism in place to service deep-dive analysis and insight requests based on these datasets. |
| DATASSENTIALS & SNAP | Datassentials provides flavour, recipe trends and prevalence on menus in the world/<br><br>MenuTrends: the world's largest menu database and gold-standard tool for trend analysis, use to generate reports on any food, ingredient or flavour and track the top or fastest-growing ingredients, flavours and dishes at restaurants today. Answers questions like "What types of healthy/functional ingredients are growing on menus today?" "What are the fastest-growing sandwich varieties at fast casual restaurants?"<br><br>Menu Adoption Cycles: The Menu Adoption Cycle (MAC) tracks any food, ingredient or other trend through its 4-stage life cycle, from Inception to Ubiquity. The MAC is Datassential's proven framework for evaluating the lifecycle of a trend. Answers questions like "What are early-stage coffee trends? What Asian dishes are beginning to trend?"<br>FLAVOR: provides demographic views of consumer preferences for any food or ingredient. This database tracks consumer awareness, trial and preference for thousands of flavours, ingredients and food items. |
| FOOD MARKETING INSTITUTE (FMI) | FMI is a food retail industry association that conducts annual consumer surveys regarding trends and habits in food retail. The focus is on Grocery and Convenience stores. |
| NATIONAL ASSOCIATION OF CONVENIENCE STORES (NACS) | NACS is a convenience store industry association that conducts annual consumer surveys free to the public. Titles include topics like "Embracing Modern Convenience: Responding to Shoppers Needs" and "Consumer Behaviour at the Pump" |
| STATISTA | The site can be overwhelming, and it might be best to use their research service where they hunt for data. Statista has a team of over 250 data analysts and data specialists that work on retrieving dat, from the 22,500 different sources that we can work with. Research requests are great if we need specific data that fits our research needs, updated figures on statistics, or clarification on data points and sources. |

**Table 3.1: Syndicated Data**

## 4.4   DATA ANALYSIS

To study the impact of COVID-19 on downstream customer behaviour, we can look at the post-event behaviour of treatment customers in the Austrian marketplace for few high value areas (HVAs) in the months of Jan-September 2018-2020. We can use YoY% growth from 2018-2019 to calculate expected 2020 indicator values, and evaluate deviations between observed and expected values of our estimations.

The key takeaway of using machine learning is that the best model should be built by experimenting with various mathematical models. This is usually done using two datasets in different stages. First the model is trained on a section of the data, training set to estimate the parameters. Then the model is tested on the rest of the data, test set or holdout dataset to get an unbiased evaluation. In case of time-series the training data and test data should be both continuous time sequences. In the end, to get a best model first the best model has to be picked by running experiments with different algorithms and then compare the model to a pre-selected baseline, if available. (*Check the appendix A for more details*)

## 5.   RESEARCH PROTOCOL

A Gantt project diagram with an estimated time of completion can be provided upon request.

# APPENDIX A

# Data Resources/ Templates

## A.1    MACHINE LEARNING

Machine learning is the study of computer algorithms that improve automatically through experience. It is seen as a subset of artificial intelligence. Related item:

## A.2    KAGGLE

Kaggle, a subsidiary of Google LLC, is an online community of data scientists and machine learning practitioners. Related item: https://www.kaggle.com/covid19

## A.3    DETECTING OUTBREAKS WITH GOOGLE SEARCHES

By Seth Stephens-Davidowitz: Mr. Stephens-Davidowitz is the author of "Everybody Lies: Big Data, New Data, and What the Internet Can Tell Us About Who We Really Are."

Stephens-Davidowitz, S. 2020. *Google Searches Can Help Us Find Emerging Covid-19 Outbreaks*. New York times, 05 April. [Online] Available: https://www.nytimes.com/2020/04/05/opinion/coronavirus-google-searches.html Accessed: 05 April 2020.

## A.4    DETECTING OUTBREAKS WITH COMMUNITY-LEVEL SURVEILLANCE

Aaron E. Carroll is a contributing opinion writer for The New York Times. He is a professor of paediatrics at Indiana University School of Medicine and the Regenstrief Institute who blogs on health research and policy at The Incidental Economist and makes videos at Healthcare Triage. He is the author of "The Bad Food Bible: How and Why to Eat Sinfully." @aaronecarroll

Aaron E. Carroll, A. 2020. *Lesson From Singapore: Why We May Need to Think Bigger*. New York times, 23 June. [Online] Available: https://www.nytimes.com/2020/04/14/upshot/coronavirus-singapore-thinking-big.html Accessed: 23 June 2020.

## A.5    UNDERSTANDING COVID IMPACT BY REGION

COVID**MINDER** reveals the regional disparities in outcomes, determinants, and mediations of the COVID-19 pandemic. Outcomes are the direct effects of COVID-19. Social and Economic

Determinants are pre-existing risk factors that impact COVID-19 outcomes. Mediations are resources and programs used to combat the pandemic.

COVID**MINDER** analysis and visualizations are by students and staff of [The Rensselaer Institute for Data Exploration and Applications](#) at [Rensselaer Polytechnic Institute](#) with generous support from the United Health Foundation. COVID**MINDER** is an open source project implemented on the [R Shiny platform](#); see the [COVIDMINDER github](#) for more information. COVID**MINDER** was directed by Kristin P. Bennett and John S. Erickson.

Mary L. Martialay, A. 2020. *Data Visualization Tool Examines Community Factors Underlying COVID-19 Outcomes*. RENSSELAER Polytechnic Institute, 13 April. [Online] Available: https://news.rpi.edu/content/2020/04/13/data-visualization-tool-examines-community-factors-underlying-covid-19-outcomes?fbclid=IwAR1B3dIABTdxoDCAXM7ShbKo_7zFuwtTxi_FTEEVSx29PElC7PQQ7qiI9lQ Accessed: 13 April 2020.

## A.6    USING NLP TO UNDERSTAND COVID

Due to the COVID-19 pandemic, scientists and researchers around the world are publishing an immense amount of new research in order to understand and combat the disease. While the volume of research is very encouraging, it can be difficult for scientists and researchers to keep up with the rapid pace of new publications. Traditional search engines can be excellent resources for finding real-time information on general COVID-19 questions like "How many COVID-19 cases are there in the United States?", but can struggle with understanding the meaning behind *research-driven* queries. Furthermore, searching through the existing corpus of COVID-19 scientific literature with traditional keyword-based approaches can make it difficult to pinpoint relevant evidence for complex queries.

Keith Hall, K. 2020. *An NLU-Powered Tool to Explore COVID-19 Scientific Literature*. Research Scientist, Natural Language Understanding, Google Research, 04 May. [Online] Available: https://ai.googleblog.com/2020/05/an-nlu-powered-tool-to-explore-covid-19.html Accessed: 04 May 2020.

## A.7    OVERVIEW OF AI APPLICATIONS

The role AI plays in the public health community's COVID-19 response In our recent virtual forum, leaders from the Federal Government, Booz Allen, and select industry and academic partners provided a candid look at how artificial intelligence (AI) is already being applied, plus explored opportunities to leverage it further.

## A.8    UNDERSTANDING THE RISK LEVEL FOR GETTING COVID-19

Guided by common values, Covid Act Now is a multidisciplinary team of technologists, epidemiologists, health experts, and public policy leaders working to provide disease intelligence and data analysis on COVID in the U.S.

We published the first version of our model on March 20. Over 10 million Americans have used the model since. We've engaged with dozens of federal, state, and local government officials, including the U.S. military and White House, to assist with response planning.

## A.9    COVID-19 EVENT RISK ASSESSMENT PLANNING TOOL

This site provides interactive context to assess the risk that one or more individuals infected with COVID-19 are present in an event of various sizes. The model is simple, intentionally so, and provided some context for the rationale to halt large gatherings in early-mid March and newly relevant context for considering when and how to re-open. Precisely because of under-testing and the risk of exposure and infection, these risk calculations provide furher support for the ongoing need for social distancing and protective measures. Such precautions are still needed even in small events, given the large number of circulating cases.

## A.10    FORMULAS

In a machine learning exercise, to get a best model first the best model has to be picked by running experiments with different algorithms and then compare the model to a pre-selected baseline, if available.

Below we have defined the evaluation metrics used to choose the best algorithm and baseline used to prove that this method will be an improved version.

## A.10.1 EVALUATION METRICS

**Mean Absolute Percentage Error (MAPE):** = *MEAN*( *ABS*( ( actual- predicted) / actual) ) * 100

**Root Mean Square Error (RMSE) =** *SQRT*( *MEAN*( *SUM*( (actual - predicted)^2)))

Run Markov Switching Auto Regression Model on growth or y-o-y growth, and get probability of COVID/ Normal periods for each time series.

Two regimes: COVID and Non-COVID

with regime transitions according to:

$$P(S_t = s_t | S_{t-1} = s_{t-1}) = \begin{bmatrix} p_{00} & p_{10} \\ p_{01} & p_{11} \end{bmatrix}$$

**saliency(term w)** = frequency(w) * [sum_t p(t | w) * log(p(t | w)/p(t))] for topics t

*src:* *see Chuang et. al (2012)*

**relevance(term w | topic t)** = λ * p(w | t) + (1 - λ) * p(w | t)/p(w)

*src:* *see Sievert & Shirley (2014*

## A.10.2 HOW THE PEAK THRESHOLDS WERE CHOSEN:

Two metrics were used to classify outlier and non-outlier forecast data. A prior *n*-week forecast moving average, and the standard deviation of this *n*-week forecast average multiplied by some scalar *k*.

If *abs(forecast(n+1) - forecastAvg(n)) > k* SD(n)* , then *forecast(n+1)* is part of a false peak forecast.

Long lasting peaks, make the COVID peak look like a non-peak, which ends up showing false peak forecasts as non-outliers. To resolve this, the false peak is adjusted downward before being included in the *n*-week moving average. *forecast(n+1)* is replaced by the average of (*forecast(n), forecast(n+1)*)

# APPENDIX B

# Towards A Supply Chain Re-Set

## B.1    TESTING RESILIENCE OF YOUR SUPPLY CHAIN: QUESTIONS TO ASK

**Strengthen the strategic view of supply**

- How responsive is my supply chain? How quickly can I make decisions?

- Do I understand my supplier reliance? Where are the critical risks?

- Where are my key suppliers located? Would diversifying geographies reduce risk?

- Do I have an inventory strategy? Does it consider the risk of future disruption?

**Invest in digital supply chain technologies**

- Do I have end-to-end visibility of my supply chain? Where are the risks and gaps?

- Can I predict demand? Where could I benefit from more powerful demand prediction?

- Do I have robust scenario planning in place? Could tools such as digital twins help me understand the potential impact on the whole business from shocks?

*Matthew Warrington and Rebecca Russell*. 2020. *Building Resilience for Australian Companies Post COVID-19*, BCG. 29 July. [Online] Available: https://www.bcg.com/en-au/capabilities/operations/building-resilience-for-australian-companies-post-covid-19 Accessed: 29 July 2020.

## B.2    IMPACT OF COVID19 ON SUPPLY CHAIN, PRODUCT ALLOCATION, TRANDING AND SEASONALITY: TOWARDS BUSNIESS RESILIENCE

Resilience is especially important today because the business environment is becoming more dynamic and unpredictable. This is a result of several enduring forces stressing and stretching business systems — from accelerated technological evolution to a greater interconnectedness of the global economy to broader issues such as rising inequality, species depletion, and climate change.

Companies can **structure** their organizations and decision processes for resilience by embracing six principles of long-lasting systems:

- **Redundancy** buffers systems against unexpected shocks, albeit at the expense of short-term efficiency. It can be created by duplicating elements (such as by having multiple factories that produce the same product) or by having different elements that achieve the same end (functional redundancy).

- **Diversity** of responses to a new stress helps ensure that systems do not fail catastrophically, albeit at the expense of the efficiencies obtainable through standardization. In business, this requires not only employing people from different backgrounds and with different cognitive profiles but also creating an environment that fosters multiple ways of thinking and doing things.

- **Modularity** allows individual elements to fail without the whole system collapsing, albeit while forgoing the efficiency of a tightly integrated organizational design. Because a modular organization can be divided into smaller chunks with well-defined interfaces, it is also more understandable and can be rewired more rapidly during a crisis.

- **Adaptability** is the ability to evolve through trial and error. It requires a certain level of variance or diversity, obtained through natural or planned experimentation, in combination with an iterative selection mechanism to scale up the ideas that work best. Processes and structures in adaptive organizations are designed for flexibility and learning rather than stability and minimal variance.

- **Prudence** involves operating on the precautionary principle that if something could plausibly happen, it eventually will. This calls for developing contingency plans and stress tests for plausible risks with significant consequences — which can be envisioned and prepared for through scenario planning, war games, monitoring early warning signals, analysing system vulnerabilities, and other techniques.

- **Embeddedness** is the alignment of a company's goals and activities with those of broader systems. It is critical to long-term success because companies are embedded in supply chains, business ecosystems, economies, societies, and natural ecosystems. Articulating a purpose — the way in which a corporation aims to serve important societal needs — is a good way to ensure that the company does not find itself in opposition to society and inviting resistance, restriction, and sanction.

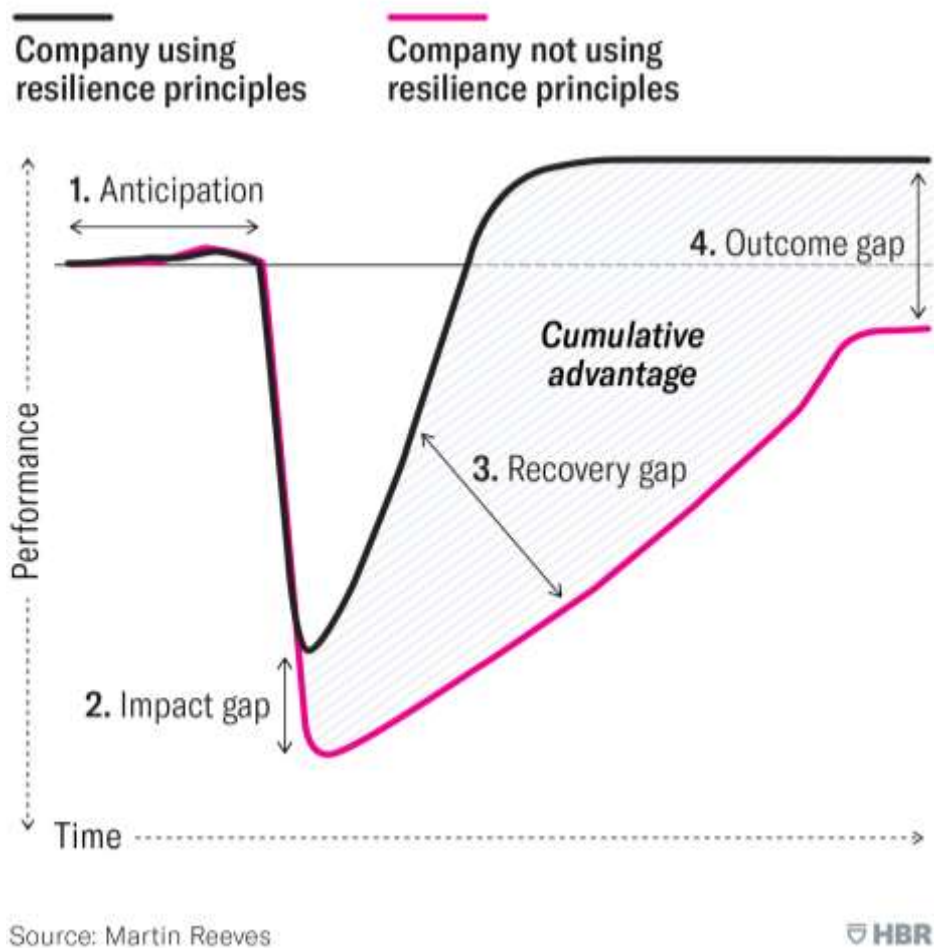

**Figure B.1: Assessing companies' relative resilience**

**B.3     HOW TO BECOME A MORE RESILIENT COMPANY**

Crises are opportunities for change. With Covid-19, companies have a unique opportunity and necessity to revisit their business models to build greater systemic resilience, starting with the following six actions.

**Seek advantage in adversity**. Don't merely endeavour to mitigate risk or damage or restore what was; rather, aim to create advantage in adversity by effectively adjusting to new realities.

**Look forward**. In the short run, a crisis many appear tactical and operational, but on longer timescales, new needs and the incapacitation of competitors create opportunities. Crises can also be the best pretext for accelerating long-term transformational change. One of the key roles for leaders is therefore to shift an organization's time horizons outward.

**Take a collaborative, systems view**. In stable times, business can be thought of as performance maximization with a given business model in a given context. Resilience, by contrast, concerns how the relationships between a business's components or between a business and its context change under stress. It requires systems thinking and systemic solutions, which in turn depend on collaboration among employees, customers, and other stakeholders.

**Measure beyond performance**. The health of a business is not captured only by measures of value extracted, which tend to be backward-looking. Measuring flexibility, adaptation, and other components of resilience is critical to building a sustainable business. This can be done quite simply by looking at either benefits or capabilities.

**Prize diversity**. Resilience depends on being able to generate alternative ways of reacting to situations, which in turn depends on the ability to see things with fresh eyes. Resilient businesses prize cognitive diversity and appreciate the value of variation and divergence.

**Change as the default**. Alibaba founder Jack Ma sees change, not stability, as the default. Resilience is less about occasional adjustments under extreme circumstances and more about building organizations and supporting systems predicated on constant change and experimentation. This is partly to avoid rigidity and partly because iterative incremental adjustment is far less risky than a massive one-shot adjustment.

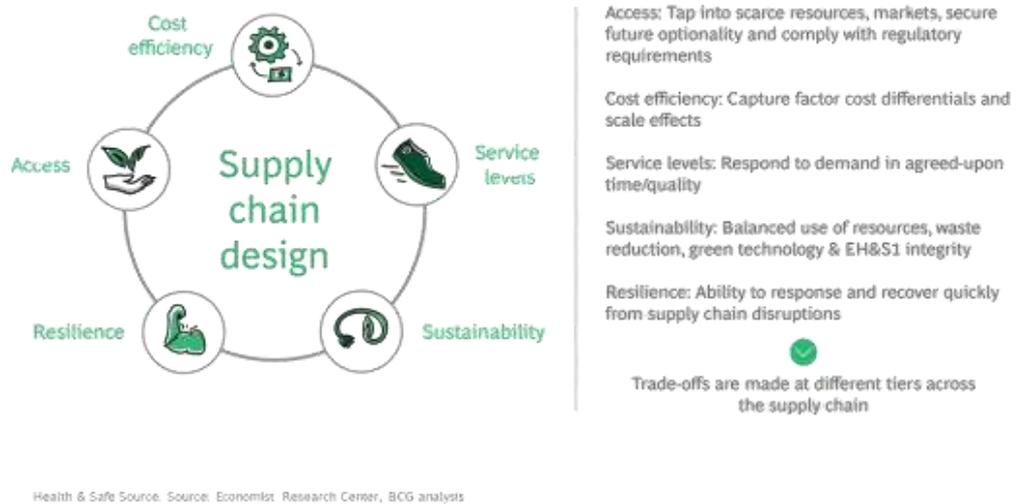

**Figure B.2: Strengthening supply chain resilience requires companies to make trade-offs**

Resilience is a supply chain design choice that senior leaders can pursue for greater stability and reduced risk. As with most design choices, resilience involves trade-offs; for example, a cost-efficient supply chain might prioritise low inventory levels, while a resilient supply chain might maintain higher levels of inventory to provide a buffer for critical items. These trade-offs may seem unnecessary when times are good, but they can be the difference between success and failure in times of crisis.

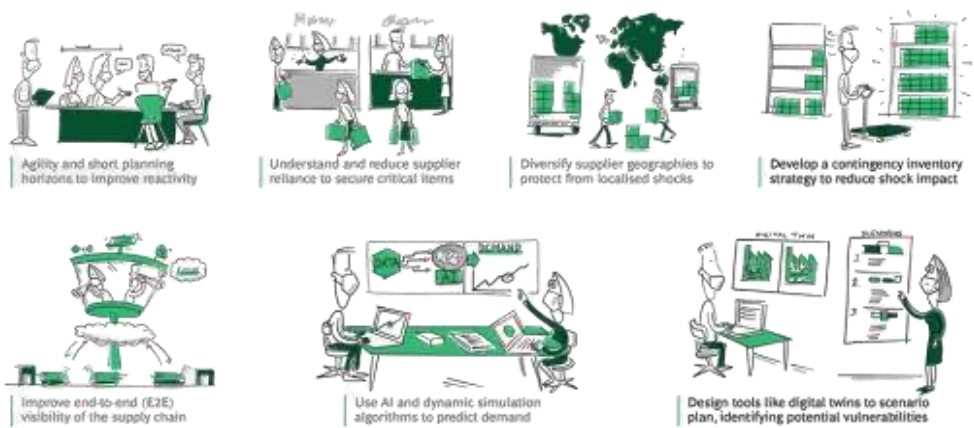

**Figure B.3: Seven levers to improve supply chain resilience**

## B.4    IMPACT OF COVID19 ON SUPPLY CHAIN, PRODUCT ALLOCATION, TRENDING AND SEASONALITY

This is the subject of our current research

# APPENDIX C

# REFERENCES

## C.1 IMPACT OF COVID-19 ON RETAIL INDUSTRY: CHANGING CUSTOMER DEMAND

In a typical global supply chain system, raw materials may be sourced from one country, the components designed in another, parts made in a third, then the pieces shipped to various assembly plants where labour costs are low. The finished products are then packaged and transported across the globe, possibly for more manufacturing or pre-sale preparation at the point of sale. As the world responds to the COVID-19 pandemic by closing large parts of their economies and their borders, the interdependence of markets has resulted in unprecedented disruptions to global supply chains. In this scenario, supply problems are inevitable, since a global network only needs one key economy or segment to malfunction to cause disruptions along the whole supply chain. The COVID-19 pandemic has thrown a spotlight on the fragility of our supply chains and on the implications this has for Australia, for example, in terms of access to vital medical personal protective equipment, chemical reagents for testing, supply of Isotopes for cancer treatments and many other critical final products and precursor elements. The consumer response to COVID 19 has seen dramatic changes to purchasing behaviour and intensified consumer demand for many products. We have witnessed panic buying, long queues and empty shelves as people rush to stock their pantries. It is important to note the current product shortages on shop shelves is not about supply but rather a product of unexpected demand and a lack of preparedness of our national to consider and manage the real risk of pandemic.

## C.2 NEED OF VIABILITY RISK MONITORING FRAMEWORK

Thinking about designing a Viability Risk Monitoring framework can help retailers anticipate and avoid more supply chain disruptions in the COVID19 world. By leveraging patented algorithms and a multitude of data sources to analyse conventional and emerging risks, this framework will isolate urgent issues and monitors your third-party network. The job is to calculate the risk each vendor presents to a retailing business—helping them improve the resiliency, efficiency, and value of their supply chain.

## C.3     NEED OF VIABILITY RISK MONITORING FRAMEWORK

From time to time, frequent as well as rare catastrophes, such as the novel coronavirus, also disrupt supply chain operations. There is a huge need of collecting and compiling data from many sources and show that there has been a marked increase in both the frequency and economic losses from natural and man-made catastrophes. The job is to find that in ncase business losses constitute a major percentage of the total losses caused by these catastrophes. The statistics suggest that for terrorist attacks, the vulnerability of U.S. business interests is much higher than others. Examination of the geographical and chronological distributions of catastrophes provides useful information for managers concerned about such disruptions. This attempt of forecasting the previsions tends to develop a catastrophe classification framework that matches different types of catastrophes to a variety of infrastructural components of supply chains. The framework also connects a variety of mitigating strategies to appropriate catastrophe types. We then identify factors that can be used to assess the vulnerability of a supply chain. They can also be useful to compare possible alternative decisions based on the vulnerability they may cause in the supply chain. To manage vulnerability in supply chains, there is a need of proposing strategies that can be implemented by a company to decrease the possibility of occurrence, provide advance warning, and cope after a disturbance.

## C.4     HOW THE COVID-19 PANDEMIC MAY CHANGE THE WORLD OF RETAILING

The world has changed dramatically in just a few months with the spread of the novel coronavirus, COVID-19. This pandemic has altered people's lives and wreaked havoc on the global economy. While the long-term effects of COVID-19 are yet to be determined, its immediate impact on retailing is significant. Retailers of essential goods such as food, groceries, and healthcare are experiencing increased demand opportunities for serving consumers at home, while facing challenges of inventory, supply chain management, delivery, and keeping their facility a safe environment. On the other hand, retailers of non-essential goods, such as apparel and footwear, are facing a significant drop in sales and are having to adopt new ways to reach and engage customers who are shopping from their home, just to sustain themselves. Some manufacturers and retailers are even pivoting and changing their product mix to suit the demand arising from the COVID-19 crisis (e.g., shoe manufacturers creating facemasks, spirit manufacturers using the same alcohol ingredient for producing and selling hand sanitizers during the present crisis).

Anne L. Roggeveen. 2020. *How the COVID-19 Pandemic May Change the World of Retailing*. [Online] Available: https://www.ncbi.nlm.nih.gov/pmc/articles/PMC7183942/ Accessed: 27 April 2020.

## C.5     CUSTOMER-INTERFACING RETAIL TECHNOLOGIES IN 2020 & BEYOND: AN INTEGRATIVE FRAMEWORK AND RESEARCH DIRECTIONS

The world of retailing is changing rapidly, and much of that change has been enabled by customer-interfacing retail technologies. This commentary offers a framework for classifying technologies, based on their primary influence on a customer's purchase journey – in the pre-purchase stage, needs management and search engagement technologies; in purchase stage, purchase transaction and physical acquisition technologies; and in the post-purchase stage, follow-up service and loyalty management technologies. We then discuss and classify forty recent retail technologies according to this framework. Finally, we identify areas that offer great potential for further research on retail technology.

## C.6 EFFECTS OF THE COVID-19 PANDEMIC ON THE GROCERY RETAIL SUPPLY CHAINS

With the characterization of the COVID-19 outbreak as a pandemic by the World Health Organization (WHO) on March 11, 2020, numerous countries fell into panic across the world trying to devise a plan of containing the virus and protecting their citizens. Due to the lack of government procedures, each country has taken different steps on how to approach the pandemic. With a mixture of social distancing measures, emphasis on proper hygiene, and warnings of how the virus can spread, the United States responded to the pandemic with the closure of non-essential businesses paired with stay-at-home orders across different states. Essential businesses that continue to operate, such as grocery stores, have been faced with increasingly large consumer demand while having to continually adapt to government health precautions and restrictions as they change daily. The goal of this paper is to identify the specific measures that grocery stores have taken to keep their businesses afloat and to establish a framework for other grocery stores and essential businesses to follow in the event of a future pandemic. This analysis includes an evaluation of consumer demand, resource availability, logistics, and the economic effect on these essential businesses.

# APPENDIX D

# Learning Resources

## D.1    CHRISALBON

Data Science & Machine Learning: To Fight for Something That Matters

Chris Albon is the Director of Machine Learning at the Wikimedia Foundation. He has spent over a decade applying statistical learning, artificial intelligence, and software engineering to political, social, and humanitarian efforts.

Related item: >> https://chrisalbon.com/#machine_learning_engineering

## D.2    VISTUAL VOCABULARY

Designing with Data (Deviation; Correlation; Ranking; Distribution; Change over time; Magnitude; Part-to-whole; Spatial; Flow)

Related item: >> https://i.redd.it/3yf4vpn5yg221.png

## D.3    MACHINE LEARNING CRASH COURSE FROM GOOGLE

Machine Learning Crash Course features a series of lessons with video lectures, real-world case studies, and hands-on practice exercises.

Related item: >> https://developers.google.com/machine-learning/crash-course/ml-intro?authuser=2

## D.4    THE ML TEST SCORE

A Rubric for ML Production Readiness and Technical Debt Reduction Eric Breck, Shanqing Cai, Eric Nielsen, Michael Salib, D. Sculley Google, Inc. ebreck, cais, nielsene, msalib, dsculley@google.com--

Related item: >>

https://storage.googleapis.com/pub-tools-public-publication-data/pdf/aad9f93b86b7addfea4c419b9100c6cdd26cacea.pdf